\documentstyle[aps,preprint]{revtex}
\begin{document}
%\preprint{NSF-ITP-97-114}
\draft

\title{\Large{Absorption and wavepackets in optically excited semiconductor superlattices driven by dc-ac fields }}

\author{W. Yan, F. Claro, and Z.Y. Zeng}

\address{Facultad de F\'isica, Pontificia Universidad de Cat\'olica de Chile, Casilla 306, Santiago 22, Chile}

\author{J. Q. Liang}

\address{Institute of Theoretical Physics, Shanxi University, Taiyuan, 030006, People's Republic of China}
%\date{\empty}
\maketitle

\begin{abstract}
 Within the one-dimensional tight-binding minibands and on-site
 Coloumbic interaction approximation,
 the absorption spectrum and coherent wavepacket time evolution in an optically
 excited semiconductor superlattice driven by
 dc-ac electric fields are investigated using the semiconductor Bloch equations.
 The dominating roles of the ratios of dc-Stark to external ac frequency, as well
as ac-Stark to external ac frequency, is emphasized. If the former
is an integer ${\cal N}$, then also ${\cal N}$ harmonics are
present within one Stark frequency, while the fractional case
leads to the formation of excitonic fractional ladders. The later
ratio determines the size and profile of the wavepacket. In the
absence of excitonic interaction it controls the maximum size
wavepackets reach within one cycle, while the interaction produces
a strong anisotropy and tends to palliate the dynamic wavepacket
localization.
\end{abstract}


\widetext

\section{Introduction}

The effect of static and/or time-dependent electric
fields on superlattices has been the subject of much recent work.
 Of particular interest is the possibility of terahertz emission caused by
 Bloch oscillations.\cite{mendez} Focus in the past has been mainly
  on the presence of static electric fields while neglecting the THz emission signal propagating
  in the superlattices. Recently however, in order to interpret the experimental
  observation of peak shifts in the Wannier-Stark absorption lines of optically excited
  superlattices, a quasi-static model has been introduced where the
  generated THz electric field is treated adiabatically.\cite{lyssenko}
 Dignam has shown, however, that this adiabatic treatment of the THz electric
 field is inadequate.\cite{dignam1} The treatment can be regarded as roughly correct only with the
 proviso that the interband inverse dephasing time is greater than the
THz field frequency. If this condition is not fulfilled it is then necessary to treat
the static and time-dependent electric fields on an equal footing.
\cite{dignam1,liuzhu1}

It is also known that the combined dc and ac electric fields can
induce many interesting and novel phenomena in superlattices.
These include the appearance of inverse Bloch
oscillators\cite{unter,kroemer}, absolute negative
conductivity\cite{keay} , negative absorption
(gain)\cite{dignam1}, and fractional Wannier-Stark
ladders\cite{zhao,cold}. In this paper we study the absorption and
wavepacket dynamics in optically excited semiconductor
superlattices under the influence of dc-ac electric fields. It is
found that the parameters from both the dc and ac electric fields
play important roles in determining their coherent dynamics
behavior. The absorption spectrum is determined
 by the ratio of the Stark frequency $\omega_B$
to the THz field frequency $\omega_{ac}$. When this ratio is an
integer ${\cal N}$, ${\cal N}$ harmonics appear within the
frequency interval $\omega_B$, while if it is a fraction of the
form $\omega_B/\omega_{ac}=p/q$ the number of harmonics is $q$,
each being identified as part of the so called fractional ladder.
It is the first time that this fractional structure is found in
the frequency domain by taking into account the excitonic
interaction. Also, the excitonic together with Fano interference
on the asymmetric lineshape of the absorption is pointed out.  On
the other hand, the extent and profile of an electron-hole
wavepacket is mainly determined by the ratio of the Stark
frequency associated with the strength of the THz field, and its
frequency. The excitonic interaction produces anisotropy in the
wavepacket profile and reduces the dynamic localization effect.

\section{Model and Numerical Results}

 The coherent dynamics of an optically excited semiconductor superlattice (SL)
driven by static and time-dependent electric fields can be
described by the following semiconductor Bloch
equations\cite{haugykoch,meier1,lindberg}

\begin{equation}
\left(\frac{\partial}{\partial t}+{\frac{e}{\hbar}\bf
F}(t)\cdot{\bigtriangledown_{{\bf k}}}\right)P_{\bf k}(t)
=-\frac{i}{\hbar}[e_{e,{\bf k}}+e_{h,{\bf k}}-i\Gamma_L]P_{\bf
k}(t)-\frac{i}{\hbar}[n_{e,{\bf k}}+n_{h,{\bf k}}-1]
\omega_{R,{\bf k}}
\end{equation}

\begin{equation}
\left(\frac{\partial}{\partial t}\pm\frac{e}{\hbar}{\bf
F}(t)\cdot{\bigtriangledown}_{\bf{ k}}\right)n_{e(h)\bf k}(t)
=-2{\rm {Im}}[\omega_{R,\bf k}P^*_{\bf k}] -\Gamma_T{n_{e,(h){\bf
k}}(t)},
\end{equation}

\noindent
where $P_{\bf k}(t)$ is the interband polarization and $n_{e,(h)\bf k}(t)$ the
electron (hole) population density in the conduction (valence) band.
The quantities $e_{i,{\bf k}}=\epsilon_{i,{\bf k}}-\sum_{\bf q}V_{|{\bf k}-{\bf q}|}n_{i,{\bf q}},
~(i=e,h)$ are the renormalized electron and hole energies due to the Coulomb interaction,
with $\epsilon_{i,{\bf k}}$ being the single-particle band energy. Also,
 $\omega_{R,\bf k}=d_{cv}{\cal E}+\sum_{{\bf k}}V_{|{\bf k}-{\bf q}|}
P_{\bf q}$ are the renormalized Rabi frequencies, with $d_{cv}$
the dipole moment and
 ${\cal E}(t)={\cal E}_0\exp(-t^2/\tau^2)\exp(-i\omega_L t)$ $(\tau=100$~fs$)$,  the
 Gaussian laser pulse profile.
 The relaxation time approximation has been assumed, with $\Gamma_L$ and
$\Gamma_T$ being the longitudinal and transverse relaxation rates, respectively.

We want to obtain the spectrally-resolved absorption coefficient
and coherent wavepacket dynamics for an optically excited SL
driven by the combined dc-ac fields $F(t)=F_0+F_1\cos(\omega_{ac}
t)$. The both components of the electric field are assumed to
 point along the same direction.

 For simplicity, a one-dimensional tight-binding (TB) approximation has been adopted
 with the two miniband widths and gap energy being $\Delta_c$
 (conduction), $\Delta_v$ (valence) and $E_g$ respectively.\cite{liuzhu1,meier1,silin}
 This approximation neglects the in-plane continuum contribution
 to the absorption, whose consequence is discussed in the following.
 Firstly, in the absence of excitonic interaction,
  the absorption $\alpha(\omega)$ is proportional to the 3-D
 joint density $\rho^{(3)}(\omega)$, which is the
 convolution product of the one dimensional joint density of states (JDOS) $\rho^{(1)}(\epsilon)$
 and 2-D continuum JDOS $\rho^{(2)}(\epsilon)$. The latter is a step function
  changing abruptly at $\epsilon=0$.
$$
\alpha(\omega)\propto\int\rho^{(2)}(\epsilon)\rho^{(1)}(\hbar\omega-\epsilon)~d\epsilon.
$$
 The one dimensional JDOS $\rho^{(1)}(\epsilon)$ looks like
 a trapezoid with the concave part in the middle, and it approximately extends
 between $[-\Delta, \Delta]$, where $\Delta$ here is the combined miniband widths
 defined as $\Delta=\Delta_c+\Delta_v$.
 Due to delayed kernel of $\rho^{(1)}(\hbar\omega-\epsilon)$ in the above expression,
 the resulting $\rho^{(3)}(\omega)$ begins to attain the finite value at $\omega=-\Delta/\hbar$.
  and becomes an approximate constant at $\omega=\Delta/\hbar$ and beyond. That is to say, that
 the behavior of $\rho^{(2)}(\epsilon)$ at $\epsilon=0$ changes completely due to
 one-dimensional JDOS. Thus, the inclusion of the in-plane continuum contribution
  can influence the height of characteristic peaks. Because JDOS is always positive,
 it can not turn the gain-type lineshape to the resonance-type shape or vice versa.
 Secondly, the contribution to the oscillator strength from  the
 in-plane continuum is negligible in the range of terahertz
 radiation of experimental interest\cite{dignam2,liuzhu2}.
 Thirdly, one also can regard the in-plane continuum as the source
 of the inhomogeneous broadening which was proposed by Thomas and von
 Plessen\cite{thomas}. This inhomegeneous broadening may smear some
 characteristics peaks appearing in our present numerical simulations.

 This TB approximation captures the qualitative features of the realistic
  superlattices, and can give a fairly good
 description of their response within the strong coupling regime.\cite{silin}
It has been used with success
 by many groups\cite{meier1,liuzhu1,liuzhu2,wxyan1} in the past.
For the Coulomb interaction we adopt the on-site model. In the
absence of the external electric fields, the contact Coloumb
 potential produces single bound states\cite{liuzhu2,egri}.
 In the dc-ac case, however, the situation changes significantly.
 For example, in the realm of the near resonance case of stark and
  ac field frequencies, unequal spacing\cite{linder}
 between the Wannier-Stark ladders due to excitonic interaction
 can make the quasienergy of the system to present more than one
 bound Floquet state. The number of bound Floquet states is
 determined by how many
 distinct deviations between excitonic Wannier-Stark ladders and
 multiple ac field quantum. This kind of Floquet
  bound states usually appears in the low Wannier-Stark ladder
  indices.
 This case of more than one bound Floquet states has been discussed in great detail by
 Liu and Zhu clearly within the picture of dressed Wannier-Stark ladders\cite{liuzhu1,liuzhu2}.

 Because we assume a {\sl weak} ultrashort laser pulse to excite the SL,
 instead of the perturbative expansion of semiconductor Bloch equations
 in terms of the order of ${\cal E}_0$,
 we take into account the full contributions from interband
 polarization. The nonlinear effect facilitated by the coupling of $n(k,t)$
 with $p(k,t)$ on the lineshape of the fractional ladders will be
 discussed in the following text.
 It is well known that the absorption coefficient can be expressed as $\alpha(\omega)
 \propto{\rm Im}\chi(\omega)={\rm Im}[{P(\omega)}/{{\cal E}(\omega)}]$, with $ P(\omega)$
 and ${\cal E}(\omega)$ being the Fourier-transformed quantities of $P(t)=\sum_kP_k(t)$
 and ${\cal E}(t)$, respectively.
 We discuss separately two cases of interest, in which the ratio of the Stark frequency
 $\omega_B=eF_0d/\hbar$ to the THz field frequency $\omega_{ac}$ is {\sl (a)} an integer ${\cal N}$,
 and {\sl (b)}, an irreducible fraction ${p/q}$.

Equations (1) and (2) were numerically integrated in the
accelerated $k$ basis, where the drift term can be eliminated. Our
results for the integer case $\omega_B=2\omega_{ac}=7\pi$ THz are
shown in Fig.1.  Figure 1(a) is for the static case, while
Fig.1(b) is for $eF_1d/\hbar\omega_{ac}=0.1$ . The values of
parameters used are: combined miniband width
$\Delta=\Delta_c+\Delta_v=40$ meV; Coulomb interaction strength
$V=$10 meV, comparable to miniband width; dephasing rates
$\Gamma_T=\Gamma_L=0.5$ TH; In the simulation of the
 coherent time evolution of the wavepackets,
the central frequency of the ultrashort laser pulse $\omega_L$ is
tuned to be in resonance with the energy-gap frequency $\omega_g$.

The Wannier-Stark resonances are clearly identified in Fig.1(a).
The frequency spacings are not equal, however, due to the
excitonic interaction\cite{linder}. In Fig.1(b) the spectrum
appears more complex, showing a weak additional structure between
any two main peaks.  These weak structures exhibit either gain or
absorption lineshapes, resulting from the nonlinear sum and
difference frequency mixing of the dc and THz electric fields. It
should be emphasized here that the ratio $edF_1/\hbar\omega_{ac}$
is a dominating factor in determining whether gain or absorption
occurs. That is to say, changing the ratio
$edF_1/\hbar\omega_{ac}$ may turn the resonance-type peaks into
gain-type dips, or vice versa. The linewidth broadening of these
structures also depends on the ratio $edF_1/\hbar\omega_{ac}$.
This can be understood by noting that the quasi-energy band
\cite{holthaus1,jauho} of a dc-ac driven single-band TB system
\cite{zak} is $\epsilon(k,\omega_{ac})=E^{(0)}+\frac{\Delta}{2}J_n
(edF_1/\hbar\omega_{ac})\cos(kd)$, with $\omega_B/\omega_{ac}=n$.
According to the generalized golden rule in the single-particle
picture the transition rate $P_{c\rightarrow v}(k)$ from the
valence dressed band to the conduction dressed band for every
component $k$, is proportional to
$\sum_{n}\delta[\epsilon_c(k,\omega_{ac})-\epsilon_v(k,\omega_{ac})+n\hbar\omega_{ac}-\hbar\omega_L)]$
. \cite{holthaus2} For every $k$, the transition can be viewed as
a two-level system (TLS) with different energy spacing. These
different energy spacings together with the $k$-{\sl mixing}
effect of the excitonic interaction lead to the broadening of the
resonance peaks and gain dips. Of course, one can select the ratio
$edF_1/\hbar\omega_{ac}$ to control the quasi-energy band widths,
hence the broadening. Another feature, apparent when comparing
Figs.1(a) and 1(b), is the relative blueshift of the {\sl main}
resonance peaks in the latter. These shifts are not equal, as
emphasized by the horizontal arrows.
  Other integer cases such as ${\cal N}$ $\omega_B/\omega_{ac}=3, ~4$ were
also calculated, yielding ${\cal N}$ resonance-type peaks or gain-type splittings in
every Stark frequency interval $\omega_B$. In order to save space, they will not be
discussed here.

Distinct asymmetric lineshapes of the peaks appear both in
dc/dc-ac field case in Fig.1. This asymmetry can be attributed to
the Fano resonance\cite{fano}. The theoretical prediction of the
Fano resonance in dc-biased semiconductor superlattices was made
by Whittaker\cite{whittaker}, Linder\cite{linder}, Glutsch, and
Bechstedt\cite{glutsch}, and was experimentally observed by
Holfeld et al\cite{holfeld}. The mechanism of the Fano resonance
is due to the coupling between the discrete and continuous
excitonic states of the relative motion, mediated by the Coulomb
  interactions\cite{whittaker,glutsch,linder,holfeld}.
 The Fano resonance in dc-ac ($\omega_B/\omega$ an integer)
 driving superlattices were studied by Liu and Zhu\cite{liuzhu2}.
 The resonance takes effect through the coupling between the
 discrete {\sl quasienergy} excitons and sidebands of their
 continua.

The other interesting case is when the ratio $\omega_B/\omega_{ac}$ equals an irreducible
 fraction $p/q$. Figure 2 shows the spectrally-resolved absorption
 coefficient for the cases $\omega_B/\omega_{ac}=1/3$, $2/3$ and $4/3$.
 Here the same excitonic interaction strength and combined miniband width
  as in Fig.1 are used.
In all cases $edF_1/\hbar\omega_{ac}=1.0$,
and $\omega_{ac}$ is fixed to be $7\pi$ THz.

Surprisingly,
the resonances appear at approximately the same frequencies in all three panels,
although the Stark frequency in the top and bottom panels differ by a factor of four.
In order to guide the eyes, dash-dot-dot vertical
lines were drawn around these resonances.
It can also be seen from the figure that some resonances split into
two peaks, which can
be understood in the following context.
 The beating frequency between the peaks in all panels is almost the
 same: $\frac{1}{3}\omega_{ac}$. This frequency equals neither the
 Stark frequency $\omega_B$, which changes from panel to panel, nor the angular frequency
 of the ac field $\omega_{ac}$.  As a matter of fact,
the quasi-energy band of the Hamiltonian driven by dc-ac electric fields with the ratio
$\omega_B/\omega_{ac}=p/q$, is the underlying mechanism for the
appearance of these peaks.
Within the one-electron and single-band approximations, these quasi-energy bands
are given by
\begin{equation}
\epsilon_n(k)=A_q(k,\omega_B)+\frac{n}{p}\hbar\omega_B,~~~~~~~~~~~~~~~~~~
n=0,1,\cdot\cdot\cdot,p-1~~,
\end{equation}
where $A_q(k,\omega_B)$ gives the band dispersion.
 \cite{zhao}
 From the last term in the right hand side of Eq.(3) it is evident that the quasi-energy band
consists of $p$ subbands with equal energy spacing
$\hbar\omega_B/p(\equiv\hbar\omega_{ac}/q)$. The absorption
spectrum reflects this energy gap through the generalized Fermi
golden rule described above, while the term $A_q(k,\omega_B)$ is
responsible for the broadening  of the spectrum. The different
values of $\omega_B$ taken in the three panels result in different
values of $A_q(k,\omega_B)$, hence, different broadening. Also,
the inclusion of the Coulomb interaction makes the positions of
the peaks to not exactly match the $\omega_{ac}/3$ rule. Yet,
since on the average there is one peak per $\omega_{ac}/3$
frequency interval, the structures are termed excitonic fractional
ladders. Their {\sl temporal} counterpart has been discussed
recently in the time-resolved four-wave mixing spectrum
\cite{wxyan1}. Results in the absence of excitonic interaction are
presented in Fig.3, where we used the same parameters as in Fig.2
except that now $V=0$. The fractional Stark ladder of spacing
$\omega_{ac}/3$ can be clearly identified, with peaks distributed
symmetrically about $\omega=\omega_g$, matching the
$\omega_{ac}/3$ rule better than the excitonic case. These have
been indicated by the vertical arrows in the upper panel of Fig.2.
Comparison of the fractional ladder lineshape in Fig.2 and Fig.3
reveals that the purely Lorentzian lineshape of some peaks in
Fig.3 is absent in the corresponding peaks in Fig.2, where the
excitonic interaction is taken into account. This demonstrates
that the excitonic interaction together with the Fano resonances
destroys the symmetry and smears some resonance peaks. Here, the
Fano interference with small parameter $q_f$ takes effect through
the coupling of the discrete excitonic fractional ladders with the
continuous sideband. It should be emphasized that the absorption
signal
 depends also on the phase difference between the ultrashort laser
 pulse and the ac field. This important effect has been found recently
 by Meier\cite{meierpe} et al in an anisotropic 3-D superlattice.

When the exciting laser field becomes strong, the nonlinear effect
due to the high order exciting laser field magnitude ${\cal E}$
begins to take effect. The typical simulation of the effect is
through the pump-probe configurations, where the weak probe laser
field is delayed to time $\tau_p$ with respect to the strong pump
field. Then the different linear and nonlinear signals can be
extracted\cite{binder,banyai}.
 This nonlinear effect makes the clear-cut fractional ladder
 harmonics broadened, tending to merge with each other. This
  is ahown in Fig.4, where the ratio of the exciting
 laser amplitude has been indicated in the upper-right corner of
 the respective panels. Here we only show the case of
 $\omega_B/\omega_{ac}=1/3$, and we use otherwise the same parameters as
  those in Fig. 3, setting the time delay $\tau_p$ to
 $50$ femtoseconds.

 There is another interesting but subtle problem when the ratio
 $\omega_B/\omega_{ac}$ is an irrational. Numerical calculation
 can not identify the essential difference between the rational
 and irrational cases. Physically, this difference will also be blurred
  by the scattering processes.

We now turn to the coherent dynamic evolution of electron-hole
wavepackets (WP). Results for a pure dc field have been previously
reported by Dignam and coworkers, where Bloch oscillations are
clearly demonstrated.\cite{dignam2} Meier et al studied relative
motion in the exciton wavefunction in {\sl momentum space} and
found a dimensional crossover due to dynamic localization induced
by a strong THz field.\cite{kenkre,meier3} More recently Hughes
and Citrin gave vivid pictures of the strongly anisotropic
excitonic WP in optically excited semiconductor quantum wells
driven by an in-plane THz field, through a time sequence of the WP
snapshots.\cite{citrin}

The WP coherent evolution may be described through the squared
modulus of the real space Fourier transform of the interband
polarization $P(z,t)=|\sum_{k} \exp(-i{ k}\cdot z)P(k,t)|^2$,
giving the real space description of the probability density of
finding an electron and hole separated by $z$ at time
$t$.\cite{citrin}  It should be noted that wavepackets can not
directly be probed by experiments, although it is theoretically
interesting. In optically excited semiconductors, the center of
mass quasi-momentum of the electron-hole is always
zero\cite{dignam1}, which makes the description of the
  behavior of the wavepackets in the center of mass difficult.

In order to elucidate the role the excitonic interaction plays on
the profile of wavepackets, we first present results in the
absence of this interaction. Snapshots at various times are shown
in Fig.5, where the Stark and ac frequencies are chosen to have
the same value $4\pi$ THz ($T_{ac}=500$ fs), and
$eF_1d/\hbar\omega_{ac}$ is set close to the first root of the
first order ordinary Bessel function (3.832).

Note that at all times the WP is symmetric about the position $z=0$.
With the elapse of time, it first splits into two parts and subsequently
into three parts at half $T_{ac}$,
 when the size of the WP acquires its largest value within a cycle. After $T_{ac}/2$, the
size begins to reduce. Due to the synchronous dynamic localization
of electron and hole, at time $t=T_{ac}$, the size of the WP experiences the severe
 shrinkage shown in the last panel of Fig.5.  At later times, the WP will undergo the
same expanding and shrinking processes in every time cycle ($T_{ac}$ ).
 This phenomenon demonstrates that dynamic localization has the ability to control the
 size of the WP. That is to say, that dynamic localization will make it
 impossible to find the electrons and holes simultaneously beyond
  a certain value of the relative coordinate. Also, comparison Fig.5
  and Fig. 6 shows that dynamic localization makes the value of
 $P(z,t)$ larger at $z=0$, demonstrating that it will enhance absorption.

Figure 6 illustrates the same case as
 that of Fig.5, except that
we choose $eF_1d/\hbar\omega_{ac}$ to be 4.5. Now the WP also
splits into three parts at time $t=T_{ac}/2$. The most important
difference, however, lies in that now the size of the WP does {\sl
not} experience shrinking towards the end of the cycle. In fact,
the size of the WP grows steadily with time, until it disappears
due to dissipation. We next turn the excitonic interaction on,
with results showing a substantial change in the profile of the
WP, as shown in Fig.7. Here we use the same parameters as in
Fig.5. The WP becomes strongly anisotropic. Its height is reduced
and although its extent still undergoes a shrinkage towards the
end of the cycle, this shrinkage is less severe as compared with
the non interacting case. The splitting structure can be viewed as
the spatial interference\cite{citrin} induced by the coulombic
re-scattering due to the oscillating motion of the photo-generated
electron-hole driven by dc-ac fields.

\section{Conclusions}

In summary, the spectrally-resolved absorption and real-time wavepackets
of an optically excited semiconductor superlattice driven by combined
parallel dc-ac electric
fields are
obtained by numerical integration of the semiconductor Bloch equations. The ratio of dc field Stark frequency $\omega_B$ to ac field frequency
$\omega_{ac}$ determines the characteristic peaks of the absorption spectrum.
When the ratio is a fraction, we identify for the first time the fractional ladders in the
energy domain by taking into account the excitonic interaction, which we take as comparable
to the miniband width. For the integer ${\cal N}$ case we find ${\cal N}$
features within the frequency interval $\omega_B$,
corresponding to gain and resonant absorption. The effect of
the dc-ac field on the size and profile of the WP shows that,
as expected, dynamic localization vigorously curtails the elongation of the WP within one cycle
 makes its evolution behavior to be of breathing mode.
The excitonic interaction
destroys the WP symmetry about its center
and softens the dynamic localization of the WP within one time cycle.

\begin{center}
{\bf ACKNOWLEDGMENT}
\end{center}
 One of the authors (W.Y.) thanks helpful discussions with Prof.
 Zdenka.
 This work was supported in part by C\'atedra Presidencial en Ciencias (F.C.) and
 FONDECYT, Grants 1990425 and 3980014.  W.Y. thanks the generous hospitality of the
 Facultad de Fisica in PUC and Departmento de Fisica in UTFSM, where part of the
 work was done. W.Y. is supported in part by the Youth Science Foundation of Shanxi
 Province of PRC. and also supported by CONICYT (3010052).

%\newpage

\begin{figure}

\caption{ The spectra-resolved absorption coefficient
$\alpha(\omega)$ for $\omega_B/\omega_{ac}=2$ is plotted as a
function of $\omega-\omega_g$ for two cases (a)
$edF_1/\hbar\omega_{ac}=0.0$ (pure dc); (b)
$edF_1/\hbar\omega_{ac}=0.1$.}
 the additional harmonics appears between two neighboring Stark resonances.
\end{figure}

\begin{figure}
\caption{The spectra-resolved $\alpha(\omega)$ for the fractional
cases $\omega_B/\omega_{ac} =1/3,2/3,4/3$ is plotted as a function
of $\omega-\omega_g$, showing the {\sl excitonic}
 fractional ladders. The ratio $edF_1/\hbar\omega_{ac}$ is chosen to be 1.0, while the other
 parameters are the same as in Fig.1}
\end{figure}

\begin{figure}
\caption {The same as Fig.2, except that the excitonic interaction is absent.
The fractional ladders
 are now symmetric
about $\omega_g$}
\end{figure}

\begin{figure}
\caption {The pump-probe spectrally-resolved four-wave mixing
signal squared $|P(\omega)|^2$ has been shown in the four panels
with the pump laser field magnitude ratio shown in the right-upper
 corner. Only the case of $\omega_{ac}=\omega_g$=1/3 is shown, the other parameters
 used here is the same as those in the Fig.2.}
\end{figure}

\begin{figure}
\caption {Coherent wave packet evolution in time in the absence of
excitonic interaction. The ratio $edF_1/\hbar\omega_{ac}$ is
chosen to be 3.832, leading to
 dynamic localization within one cycle.}
\end{figure}

\begin{figure}
\caption {The same as Fig.4, except that the ratio $edF_1/\hbar\omega_{ac}$ is chosen to be 4.50,
  leading to absence of dynamic localization}
\end{figure}
\begin{figure}
\caption {The same as Fig.4, except that excitonic interaction is taken into account}
\end{figure}

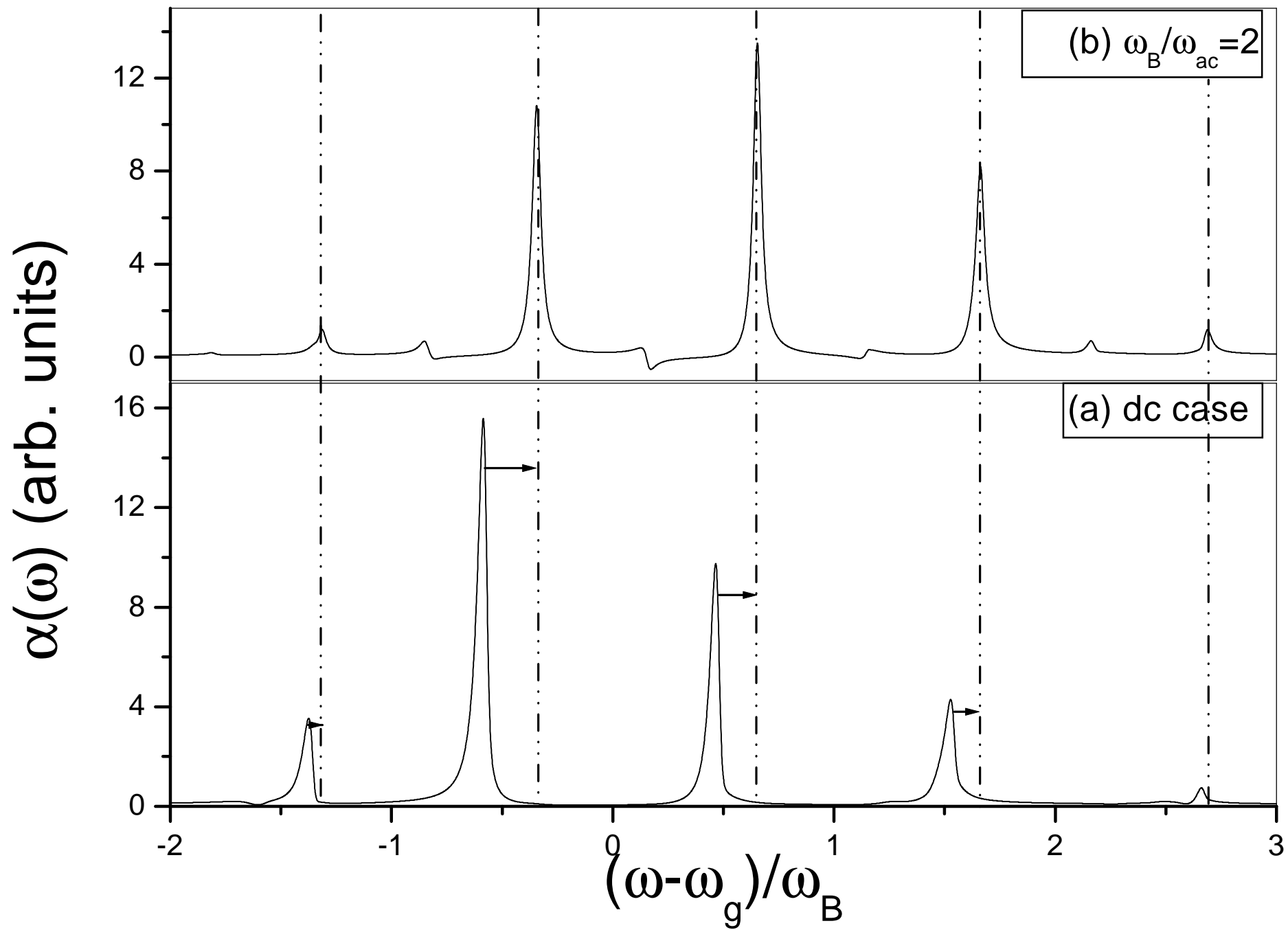
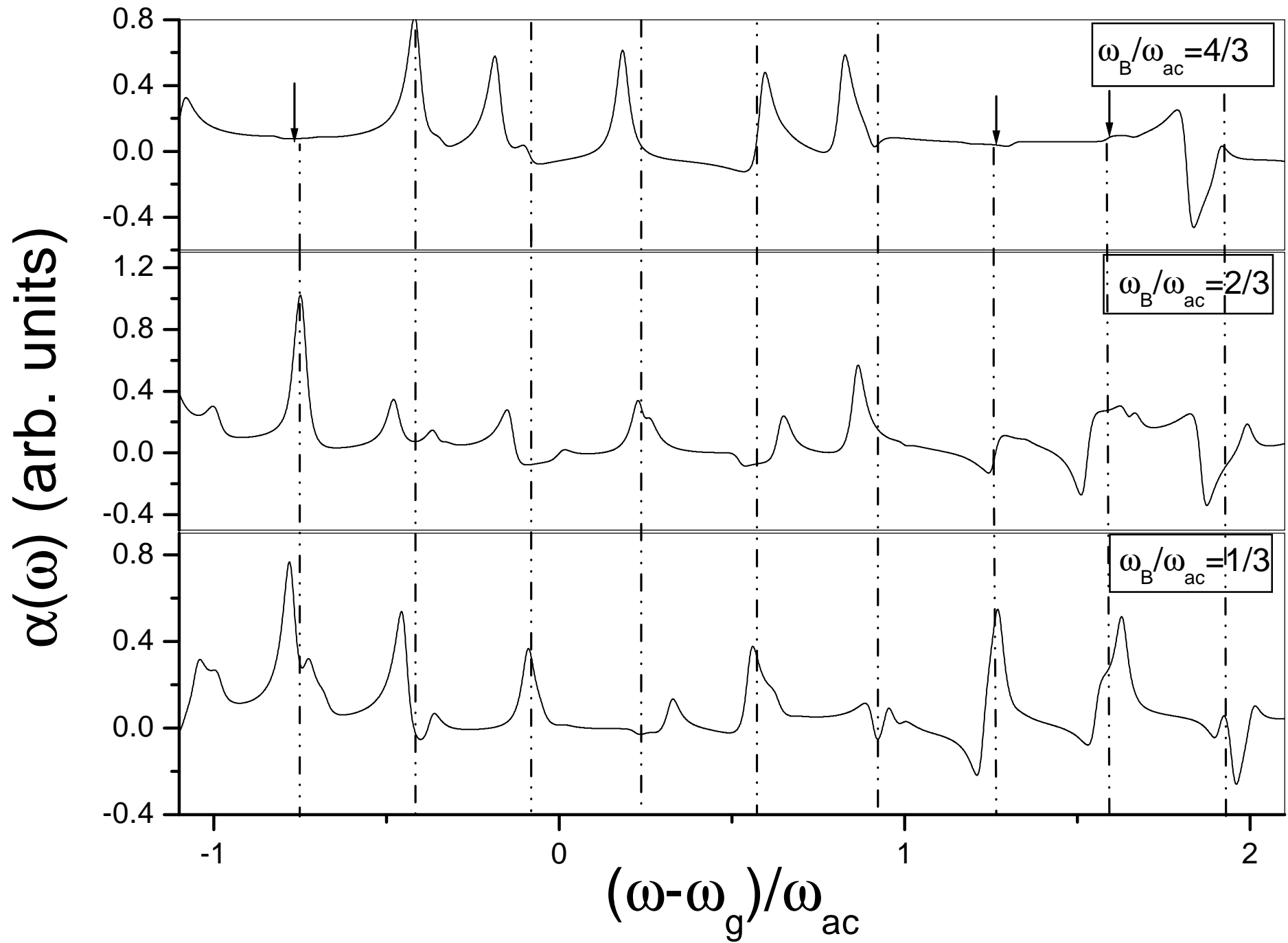
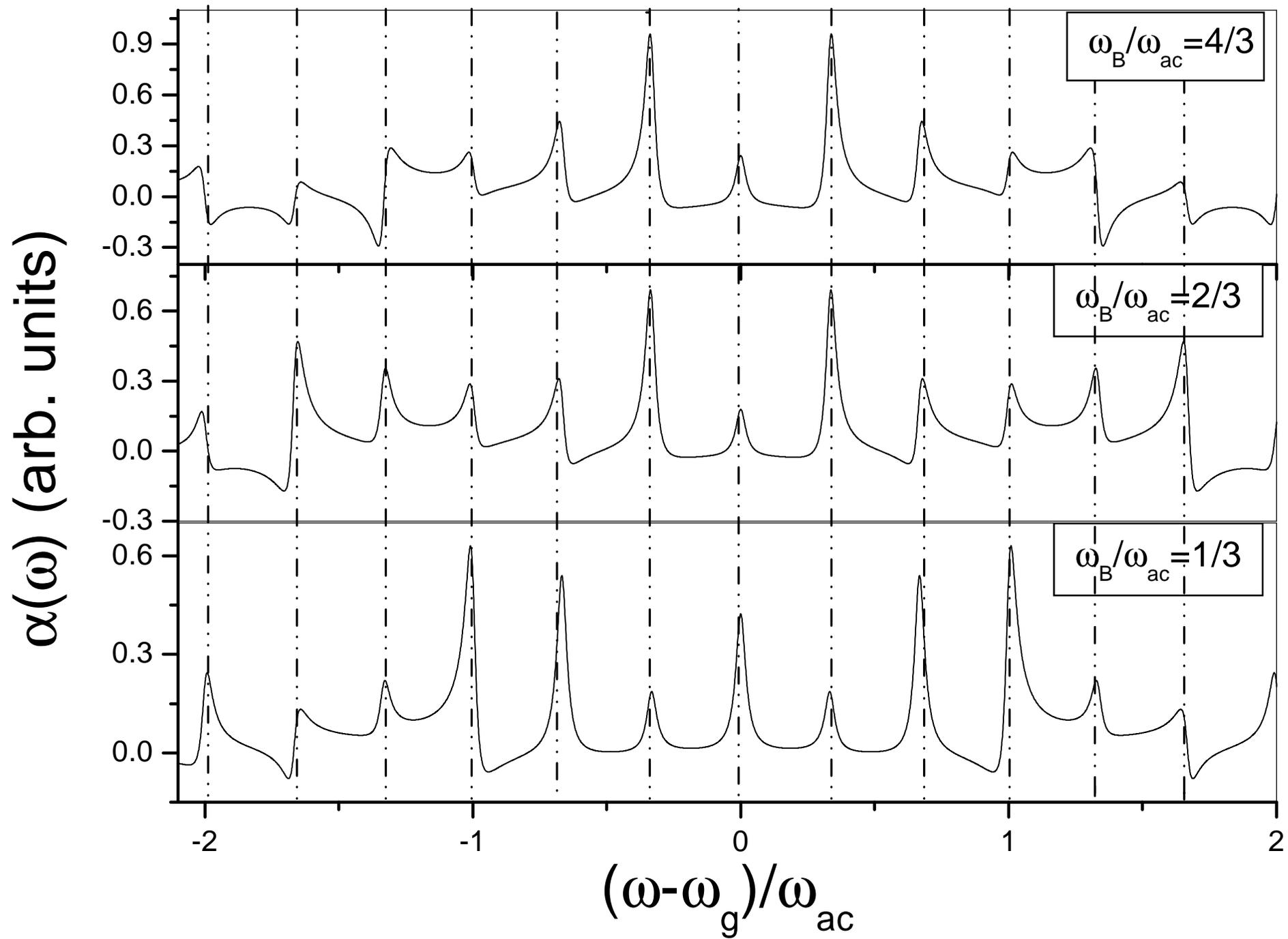
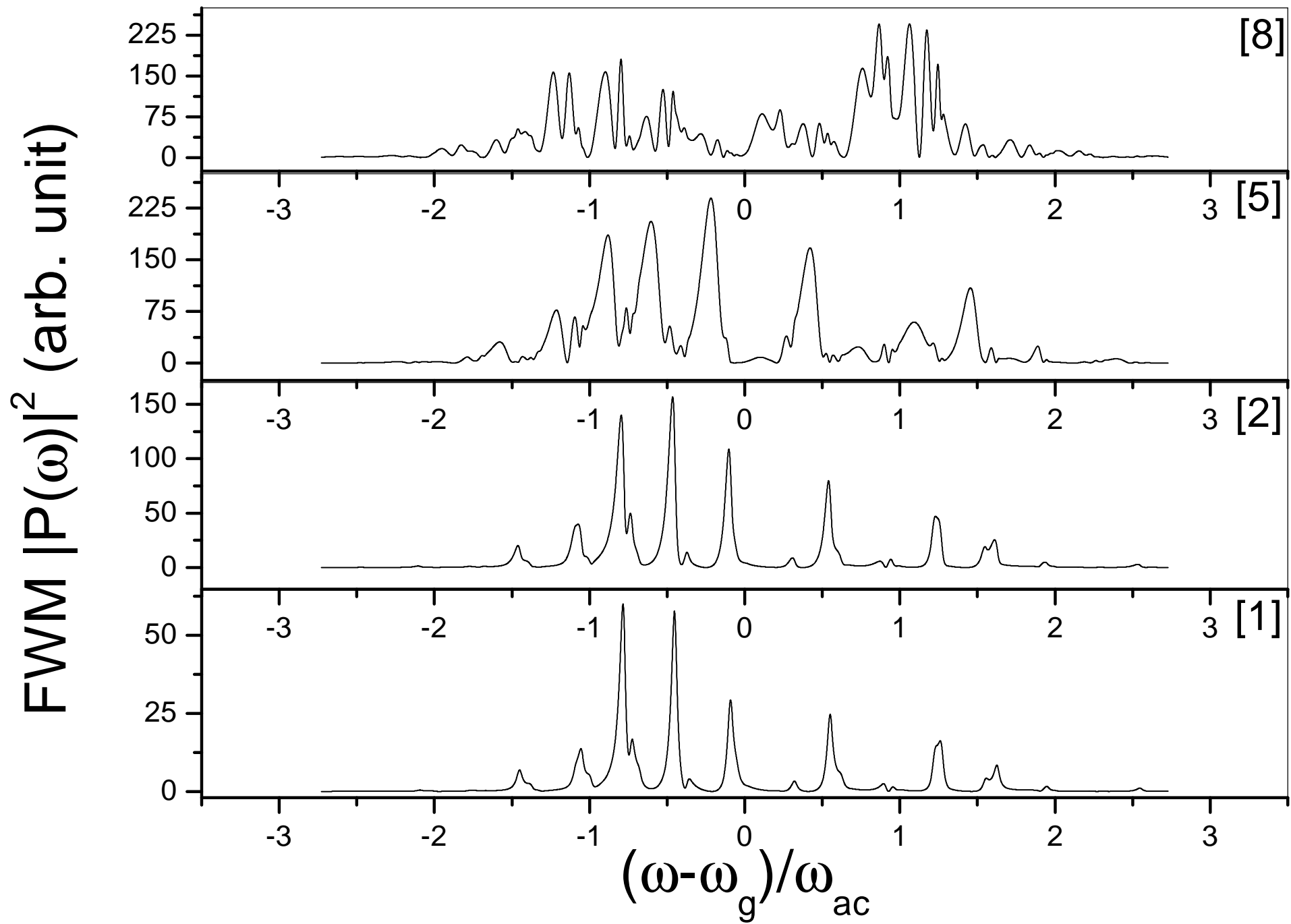
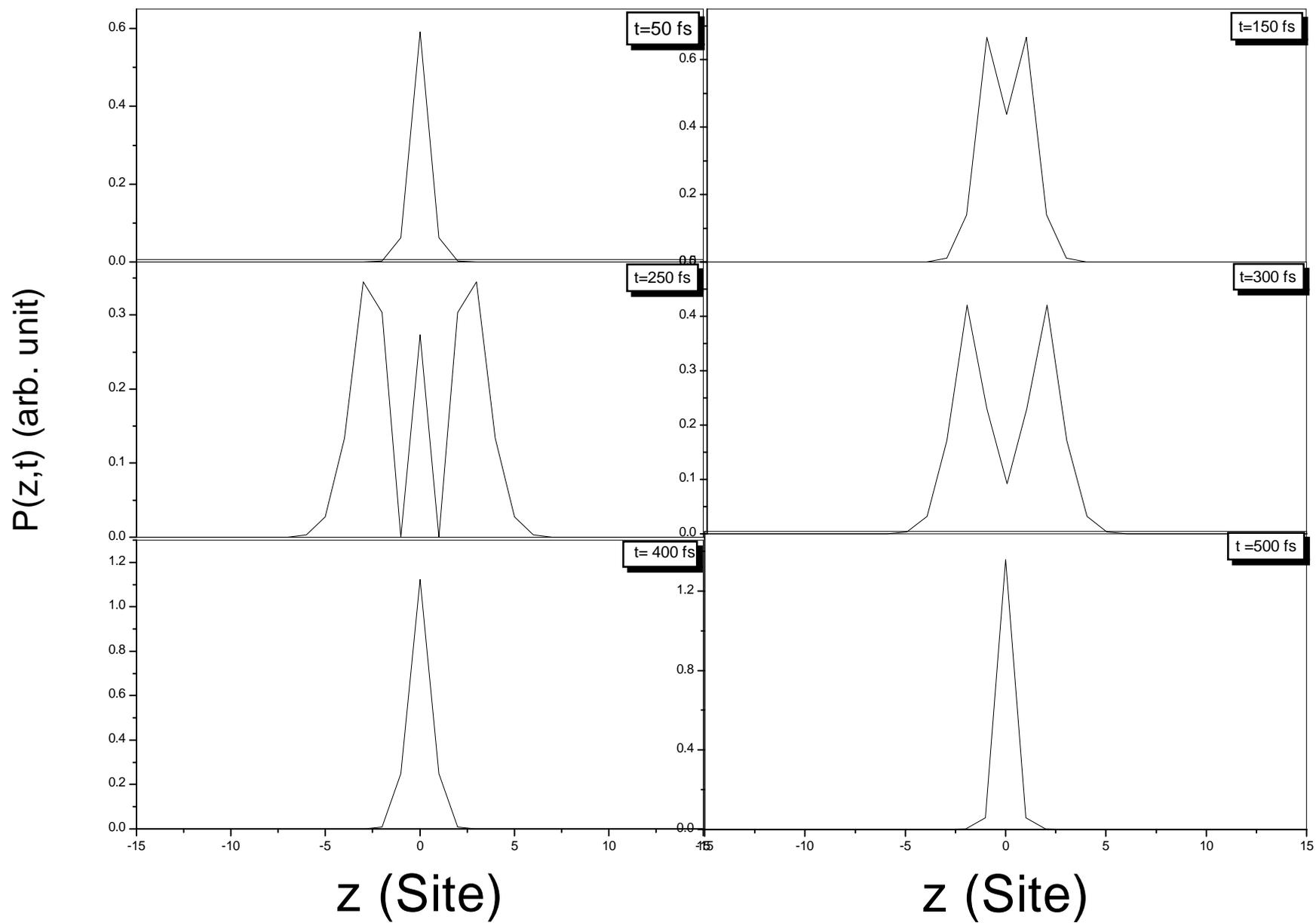
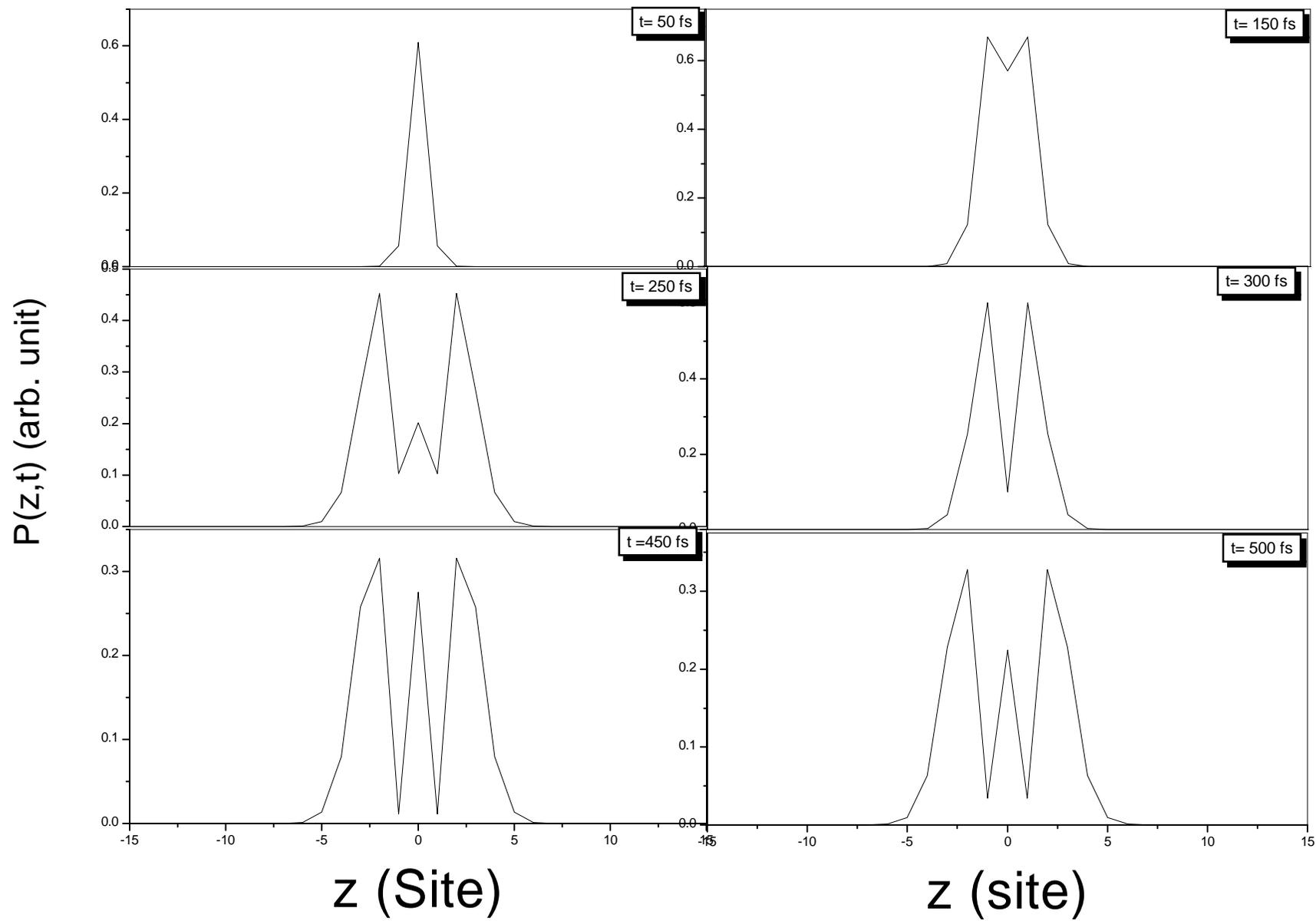
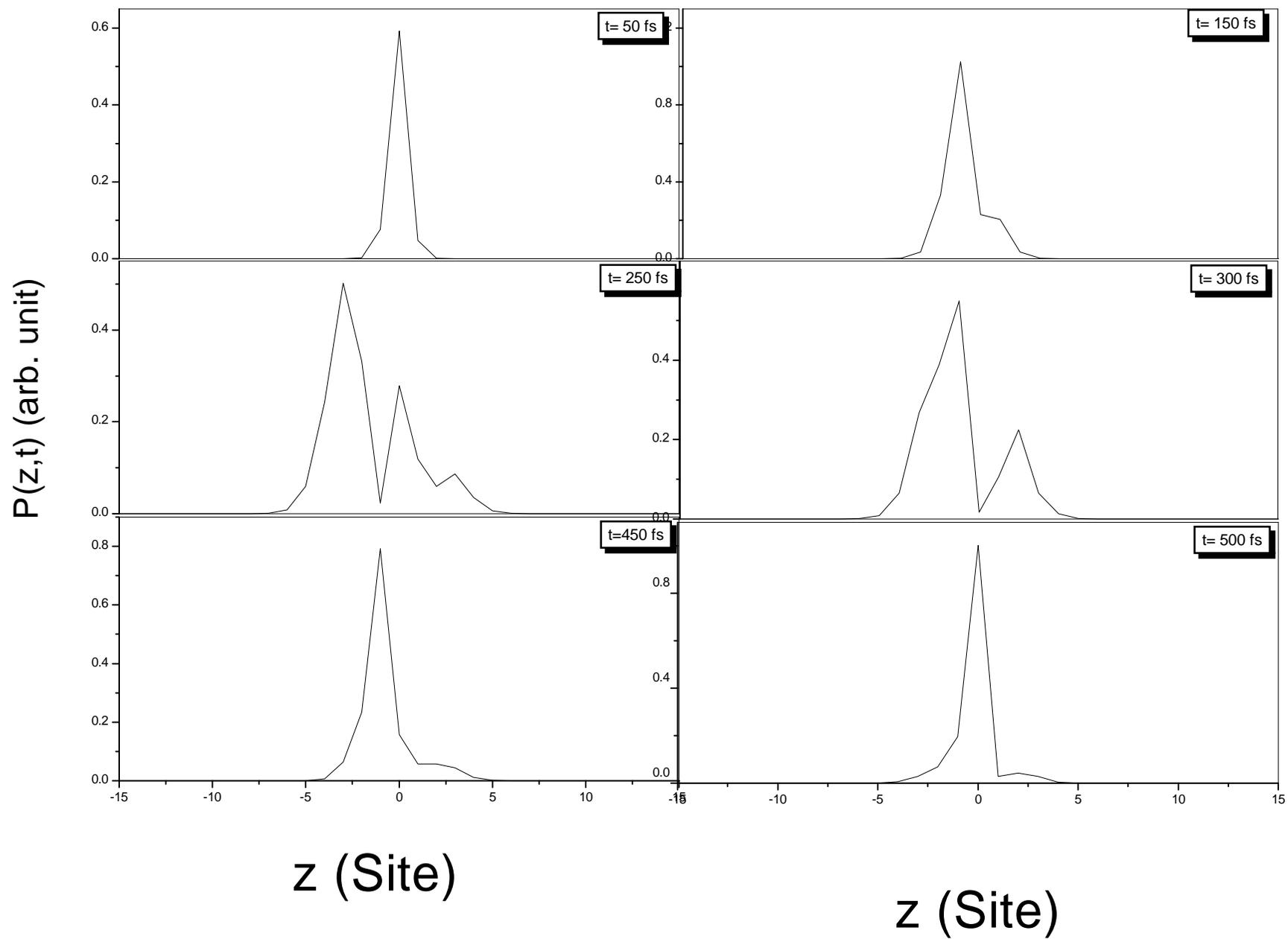
\end{document}